\documentclass[sigconf]{acmart}

\AtBeginDocument{%
  \providecommand\BibTeX{{%
    \normalfont B\kern-0.5em{\scshape i\kern-0.25em b}\kern-0.8em\TeX}}}

\usepackage{algorithm}
\usepackage{algorithmic}
\usepackage{multirow}
\usepackage{bbding}
\usepackage{algorithm}
\usepackage{algorithmic}

\usepackage{makecell}
\usepackage{bigstrut}

\usepackage{graphicx}
\usepackage{subfigure} 
\usepackage{hyperref }

\usepackage{multirow,multicol}
\usepackage[most]{tcolorbox}

\usepackage{utfsym}
\usepackage{soul}

\usepackage[flushleft]{threeparttable}
\usepackage{xspace}
\usepackage{pifont}
\usepackage{enumitem}
\usepackage{amsmath}%
\usepackage{MnSymbol}%
\usepackage{wasysym}%
\usepackage{pbox}
\usepackage{indentfirst}
\usepackage{color}

\usepackage{listings}
\makeatletter
\DeclareRobustCommand\onedot{\futurelet\@let@token\@onedot}
\def\@onedot{\ifx\@let@token.\else.\null\fi\xspace}
\def\eg{\emph{e.g}\onedot} 
\def\ie{\emph{i.e}\onedot}

\makeatother

\setcopyright{acmcopyright}
\copyrightyear{2018}
\acmYear{2018}
\acmDOI{XXXXXXX.XXXXXXX}

\acmPrice{15.00}
\acmISBN{978-1-4503-XXXX-X/18/06}

\begin{document}

\title{Large Language Model-Aware In-Context Learning for Code Generation}

\author{Jia Li}
\affiliation{%
  \institution{Key Lab of HCST (PKU), MOE, SCS, Peking University
}
  \city{Beijing}
  \country{China}
}
\email{lijiaa@pku.edu.cn}

\author{Ge Li}
\affiliation{%
  \institution{Key Lab of HCST (PKU), MOE, SCS, Peking University}
  \city{Beijing}
  \country{China}
  }
\email{lige@pku.edu.cn}

\author{Chongyang Tao}
\affiliation{%
  \institution{Peking University}
  \city{Beijing}
  \country{China}
  }
\email{chongyangtao@gmail.com}

\author{Jia Li \male, Huangzhao Zhang}
\affiliation{%
  \institution{Key Lab of HCST (PKU), MOE, SCS, Peking University}
  \city{Beijing}
  \country{China}
  }
\email{lijia@stu.pku.edu.cn}
\email{zhang\_hz@pku.edu.cn}

\author{Fang Liu}
\affiliation{%
  \institution{Beihang University}
  \city{Beijing}
  \country{China}
  }
\email{fangliu@buaa.edu.cn}

\author{Zhi Jin}
\affiliation{%
  \institution{Key Lab of HCST (PKU), MOE, SCS, Peking University}
  \city{Beijing}
  \country{China}}
\email{zhijin@pku.edu.cn}







\renewcommand{\shortauthors}{Li, et al.}

\begin{abstract}



Large language models (LLMs) have shown impressive in-context learning (ICL) ability in code generation. LLMs take a prompt consisting of requirement-code examples and a new requirement as input, and output new programs without any parameter updates. Existing studies have found that the performance of ICL is highly dominated by the quality of selected examples and thus arises research on example selection: given a new requirement, several examples are retrieved from a candidate pool for ICL. However, existing approaches are mostly based on heuristics. They randomly select examples or only consider the textual similarity of requirements to retrieve, leading to sub-optimal performance. In this paper, we propose a novel learning-based selection approach named LAIL (\underline{L}LM-\underline{A}ware \underline{I}n-context \underline{L}earning) for code generation. Given a candidate example, we exploit LLMs themselves to estimate it by considering the generation probabilities of ground-truth programs given a requirement and the example. We then label candidate examples as positive or negative through the probability feedback. Based on the labeled data, we import a contrastive learning objective to train an effective retriever that acquires the preference of LLMs in code generation. We argue that considering the feedback from LLMs themselves is logical and our approach can select suitable examples for ICL since it is consistent with the fact that LLMs have different prior knowledge and preferences. To evaluate our approach, we apply LAIL to three LLMs and evaluate it on three representative datasets (\eg, MBJP, MBPP, and MBCPP). Extensive experiments demonstrate that LATA outperforms the state-of-the-art baselines by 11.58\%, 6.89\%, and 5.07\% on CodeGen, and 4.38\%, 2.85\%, and 2.74\% on GPT-3.5 in terms of Pass@1, respectively. Human evaluation further verifies the superiority of our approach in three aspects. We also find our LAIL has surprising transferability across LLMs and datasets.

\end{abstract}



\maketitle

\section{introduction}
\label{sec:intro}

\begin{figure*}[t]
\centering
\includegraphics[width=0.98\linewidth]{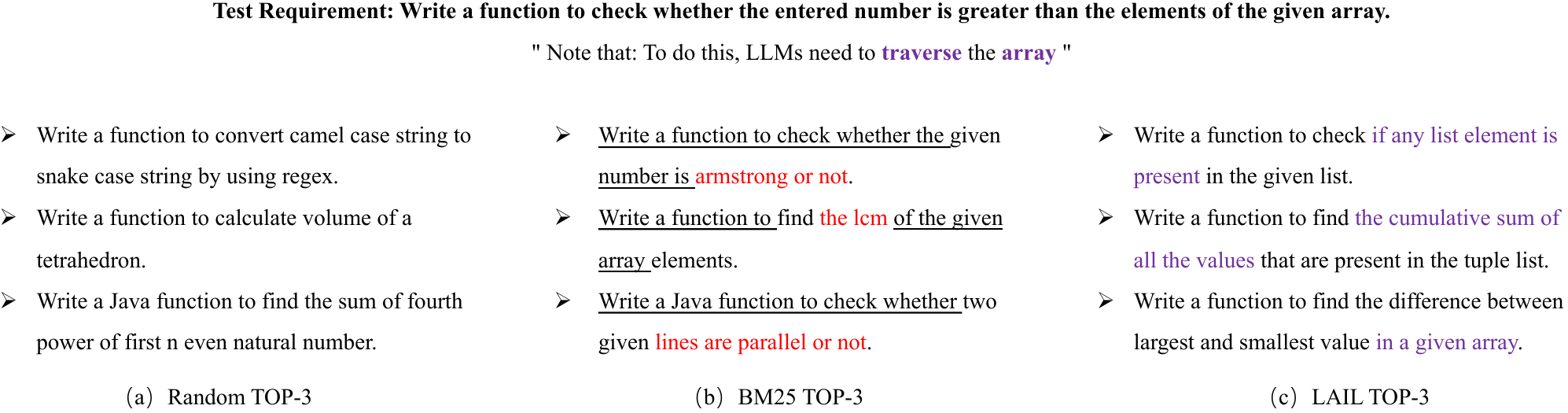}
\caption{Exhibition of the selected top-3 examples by random, BM25, and our LAIL approaches.}
\label{fig: motivation example}
\end{figure*}

Code generation aims to automatically generate the source code given a natural language requirement \cite{SkCoder, Tree_Code_Generation, bigcode_survey}.
It plays an important role in improving the productivity of software development. 
With the emergence of large language models (LLMs), in-context learning (ICL) \cite{brown2020language} becomes a prevailing paradigm and achieves impressive performance in code generation \cite{chen2023teaching, AlphaCode, li2023towards}. Given limited examples as the prompt, ICL imitates the human ability to leverage prior knowledge to generate programs without parameter updates. However, ICL comes along with the robustness problem \cite{liu2021makes}, which is sensitive to the selected examples in prompts, resulting in the performance usually varying from almost random to near state-of-the-art performance.

Despite the importance of selecting examples, only a few studies attempt to investigate example selection in code generation \cite{chen2021evaluating, bareiss2022code, li2023towards}.
One line of work is to randomly select examples from a candidate pool \cite{chen2021evaluating, jiang2023self}. The selected examples are usually improper to the test requirement and their semantic distributions vary a lot, resulting in unstable performance. 
Another line is based on learning-free heuristics \cite{li2023towards}, which leverages off-the-shelf retrievers such as BM25 \cite{robertson2009probabilistic} to select examples. 
They calculate the textual similarity between a test requirement and the requirement of examples, then select candidate items with high similarities. Despite the improved performance, these approaches only consider textual matching among requirements and thus are easily biased by exterior lexicon features. 
Figure \ref{fig: motivation example} reports the top-3 examples by random selection and the BM25 approach for the test requirement ``write a function to check whether the entered number is greater than the elements of the given array'' in MBPP dataset \cite{austin2021program}. To accomplish this requirement, LLMs need to perform traversal operations in an array.
Randomly selected examples are irrelevant to the test requirement. 
Although retrieved examples by BM25 have a large amount of textual overlap with the test item labeled by underline, the overlapping words are trivial such as `` write a function''. Meanwhile, the operations of non-overlapping words in red color, such as ``lines are parallel or not'', can provide less information to LLMs.
Examples in ICL are supposed to provide information that LLMs require, and then guide them to generate programs, but the existing approaches leave a huge gap. They do not consider the prior knowledge in LLMs and ignore their preferences. Although one may remedy it by inputting more examples in the prompt, it is highly impractical due to the limited input length of LLMs.
Therefore, selecting applicable examples for ICL becomes more urgent. 

In this paper, we propose a novel learning-based approach dubbed LAIL to select in-context examples.  
Instead of selecting examples via textual similarity, given a candidate example, our approach leverages LLMs themselves (LLM-aware) to measure it by incorporating the prediction probability of ground truth given a requirement and the candidate item. To quantify the probability feedback, we design a new metric and label candidate examples based on their metric scores. We treat examples with higher metric scores as positive examples, meanwhile, label examples equipped with lower scores as negative examples since good examples should be beneficial for LLMs to generate correct programs.
Based on labeled data, we train a neural retriever through a contrastive learning objective to acquire the preference of LLMs in code generation.
From Figure \ref{fig: motivation example}, we can find that although not textually similar to the test requirement, the selected examples of LAIL contain the ``traversal'' operation and are related to the array as shown in purple color tokens.


We evaluate LAIL on three representative LLMs including ChatGPT \cite{ChatGPT}, GPT-3.5 \cite{GPT}, and CodeGen (16B) \cite{zhong2022codegen}. We conduct extensive experiments on three datasets (\ie, MBJP (Java) \cite{athiwaratkun2022multi}, MBPP (Python) \cite{austin2021program}, and MBCPP (C++) \cite{athiwaratkun2022multi}). We use a widely used evaluation metric  Pass@k (k = 1, 3, 5) to measure the performance of different approaches. We obtain some findings from experimental results. 
\ding{182} In terms of Pass@1, LAIL significantly outperforms the state-of-the-art (SOTA) baselines by 11.58\%, 6.89\%, and 5.07\% on CodeGen, and 4.38\%, 2.85\%, and 2.74\% on GPT-3.5, respectively.
\ding{183} We conduct a human evaluation to measure generated programs in three aspects (\eg code correctness, quality, and maintainability) and further prove the superiority of LAIL. 
\ding{184} Our approach has surprising transferability across LLMs and datasets and brings satisfying improvements to them.
\ding{185} We investigate two plausible choices to estimate candidate examples by LLMs themselves and demonstrate the effectiveness of our design. 


We summarize our contributions in this paper as follows:
\begin{itemize}[topsep=0pt,noitemsep] 
\setlength{\itemsep}{1pt}
\setlength{\parskip}{0pt}
\setlength{\parsep}{-1pt}
\setlength{\leftmargin}{-1pt}

\item[$\bullet$] Our paper investigates example selection for ICL and argues that a good approach should consider what knowledge LLMs themselves require. 

\item[$\bullet$] We propose a novel learning-based approach dubbed LAIL. Given an example, LAIL exploits LLMs themselves to estimate it by calculating the generation probabilities of ground truth given a requirement and the example. We then label candidate examples via their probabilities. Based on labeled data, we import contrastive learning to optimize a retriever. 


\item[$\bullet$]  We evaluate our approach on three LLMs and conduct extensive experiments on three datasets. Qualitative and quantitative experiments reveal that LAIL significantly outperforms the state-of-the-art baselines.

\end{itemize}

\section{Background}
\label{sec:related}

\subsection{Large Language Models}

In this section, we focus on large language models (LLMs) for source code. LLMs are neural networks that aim to learn the statistical patterns in programming languages \cite{allamanis2018survey}. LLMs are pre-trained on a large-scale unlabeled code corpus with the next tokens prediction objective. Specifically, given a program with the token sequence $C = \left\{c_1, c_2, \cdots, c_n  \right\}$, LLMs are trained to predict the next token based on some previous tokens:

\begin{equation}\label{api}
\begin{aligned}
\mathcal{L}_{NTP} (C) = - \sum_{i}\texttt{log} P(c_i|c_{i-j}, \cdots, c_{i-1}; \Theta)
\end{aligned}
\end{equation}
where $j$ is the window length of previous tokens and $\Theta$ means parameters of the LLM.

After being pre-trained, LLMs are adapted to a specific downstream task. At the beginning stage, LLMs are used in a fine-tuning manner, which are continually optimized on specific code generation datasets. With the size of LLMs growing rapidly such as ChatGPT \cite{ChatGPT} and CodeGen \cite{zhong2022codegen}, fine-tuning is neither economical nor practical, in contrast, a convenient solution ICL arises.

\subsection{In-Context Learning}

In-context Learning (ICL) refers to an emerging ability of LLMs. 
Formally, given a set of requirement-code examples $T = \left\{x_k, y_k\right\}_{k=1}^m$, a test requirement $x_{t}$ and a LLM with frozen parameters $\Theta$, ICL defines the generation of a program $y_{t}$ as follows:
\begin{equation} \label{icl-formula}
\begin{aligned}
y_{t} \sim P(y_{t} |\underbrace{x_1, y_1, \cdots, x_m, y_m}_{context}, x_{t}, \Theta)
\end{aligned}
\end{equation}
where $\sim$ represents decoding strategies such as greedy decoding and nuclear sampling \cite{holtzman2019curious} in code generation.
The generation procedure is attractive as the parameters of LLMs are not need to be updated when executing a new task, which is efficient and practical.

As demonstrated in Formula \ref{icl-formula},
LLMs learn the task and generate programs depending on the selected examples. The performance of ICL can vary from almost random to near the state-of-the-art approach due to the different qualities of in-context examples. Thus, selecting appropriate examples is a significant research topic. In this paper, we propose a novel learning-based approach to select in-context examples for code generation.

\section{Method: LAIL}
\label{sec:method}

\begin{figure}[t]
\centering
\includegraphics[width=0.8\linewidth]{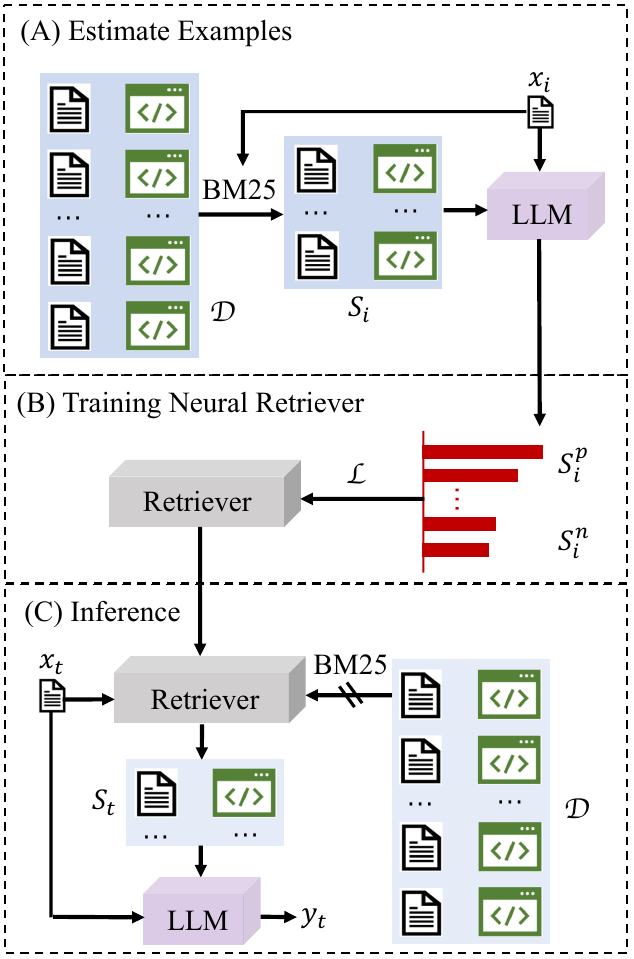}
\caption{The overview of our LAIL. LAIL use LLMs themselves to estimate candidate examples and label them as positive and negative (A). Based on the label date, LAIL then trains a retriever to align with the preference of LLMs with a contrastive loss (B). Given a test requirement, the optimized retriever selects several examples as a prompt that is inputted to LLMs for code generation (C).}
\label{fig:workflow}
\end{figure}

In this section, we propose an effective learning-based approach LAIL to select examples for ICL in code generation, as shown in Figure \ref{fig:workflow}. 
For a candidate example, LAIL exploits LLMs themselves to estimate it  by incorporating the prediction probability of the ground-truth program given a requirement and the example, then labels examples as positive and negative according to their probability feedback (Section \ref{3.1}). Our approach imports a contrastive learning objective to optimize a neural retriever, resulting in alignment with the preference of LLMs (Section \ref{3.2}). In the inference stage, given a test requirement, LAIL selects a set of examples as a prompt and a LLM generates programs with the prompt (Section \ref{3.3}).



\subsection{Estimate and Label Examples }
\label{3.1}




Different LLMs are equipped with diverse prior knowledge, thus they have disparate preferences for candidate examples in code generation. To mitigate this phenomenon, we argue that a good selection approach should align with the preference of LLMs. 
Given a candidate example, we exploit LLMs themselves (LLM-aware) to measure it by incorporating the prediction probability of the ground-truth program given a requirement and the candidate item. Based on the probability feedback, we label candidate examples as positive and negative that are then used to train the retriever in Section \ref{3.2}.

Formally, given a requirement in the candidate pool, we use LLMs themselves to estimate whether each other candidate in the pool is beneficial to generate the ground truth of the requirement, and label each candidate based on its beneficial degree. 
Following previous studies \cite{li2023towards, dong2022survey}, this paper uses the training set as the candidate pool. 
The goal of this procedure can be formulated as follows:
\begin{equation}
\begin{aligned}
\mathcal{D} = \left\{ (e_i, \left\{S_{i}^p, S_{i}^n \right\}) \right\}_{i=1}^{N}
\end{aligned} 
\end{equation}
\label{ICL}
where $e_i = (x_i, y_i)$ is the $i$-th examples in the training set. $S_{i}^p$ and $S_{i}^n$ are the positive and negative example set of $e_i$, respectively. $N$ is the number of examples in the training set $\mathcal{R}$.

Modeling on the full space of the training set is quadratic and thus prohibitive. To mitigate this limitation, we introduce a two-stage process to estimate examples in the training set.
In the first stage, for each example $e_i$, we use an off-the-shelf retriever to select a set of examples $S_i = \left\{(x_q^i, y_q^i) \right\}_{q=1}^{t}$ from $\mathcal{R}$ where $t$ $\ll$ $N$. In this paper, we apply BM25 \footnote{We also use random and semantics-based approaches to filter examples and find that BM25 is more effective and efficient.}  \cite{robertson2009probabilistic} to construct $S_i$, which measures textual n-gram matching between $x_i$ and $x_q^i$. According to BM25 scores, the top-$r$ examples with higher BM25 scores constitute the set $S_i$. 
In the second stage, we leverage LLMs themselves to estimate each example in $S_i$. Concretely, we feed each example $(x_q^i, y_q^i)$ in $S_i$ and $x_i$ into the LLM, and acquire the prediction probability of ground-truth program $y_i$. To quantify LLMs' probability feedback, we design a metric $\mathcal{M}$ which is defined as follows:
\begin{equation}
\begin{aligned}
\mathcal{M} = \frac{1}{|y_i|} \times P_{LLM} (y_i | x_q^i, y_q^i, x_i)
\end{aligned} 
\end{equation}
\label{ICL}
\begin{equation}
\begin{aligned}
P_{LLM} (y_i | x_q^i, y_q^i, x_i) = \sum_{u=1}^{|y_i|} log(p (t_{i,u}|x_q^i, y_q^i, x_i, t_{i,<u}))
\end{aligned} 
\end{equation}
\label{ICL}
where $y_i=\left\{t_1, \cdots, t_b \right\}$ and $t_u$ is the $u$-th token in $y_i$. The metric can reflect how helpful the example $(x_q^i, y_q^i)$ is for generating the target program $y_i$, which represents the preference of the LLM.

We then rank all examples in the set $S_i$ according to their metric scores $\mathcal{M}$ from high to low. The top-$z$ examples with higher scores are inputted into the positive example set $S_{i}^{p}$, meanwhile, the bottom-$v$ examples constitute the negative example set $S_{i}^{n}$ since good examples should be beneficial for LLMs to generate correct programs.
We apply this two-stage procedure to the entire training set and finally acquire the labeled data $\mathcal{D}$.

In our labeled data, the more helpful examples are labeled as positive examples and the examples with lower metric scores are treated as negative examples, which can reflect the preference of LLMs to candidate examples given a specific requirement.

\subsection{Training Neural Retriever}
\label{3.2}



As described in Section \ref{3.1}, the labeled data can reflect the preference of LLMs.
In this section, we use the labeled data $\mathcal{D}$ to train a neural retriever, aiming to align with the preference of LLMs. After being trained, given a test requirement, the retriever can select a set of candidate examples as a prompt that is beneficial for LLMs to generate correct programs. 

To optimize the retriever, we import a contrastive learning objective. The objective targets to draw a test requirement with helpful training examples together and push the given requirement with knowledge-limited examples apart. Formally, for an example $e_i$ in the labeled data $\mathcal{D}$, we randomly select an item $e_i^p$ from $S_{i}^p$ as its corresponding positive example. Meanwhile, we randomly pick a training example $e_i^n$ from the set $S_{i}^n$ and select $h$ examples from the set $\mathcal{H}$ (\eg, $\mathcal{H} = \mathcal{R} \setminus S_{i}$) as its negative example set $\mathbb{N}_{i}$. 
Next, We apply GraphCodeBERT \cite{guo2020graphcodebert} to encode their requirements and acquire corresponding representations. Then, we model the relations of these requirements with contrastive loss. Following SimCLR \cite{chen2020simple}, the learning objective $\mathcal{L}$ is formulated as:
\begin{equation}
\begin{aligned}
\mathcal{L} = & - \bigg[log \frac{e^{s(E^i_{[CLS]}, E^{i, p}_{[CLS]})/\tau}}{\sum_{E^{i,n}_{[CLS]} \in \mathbb{N}_{i}}e^{s(E^i_{[CLS]}, E^{i,n}_{[CLS]})/\tau}} \bigg]
\end{aligned}
\end{equation}
where $\tau$ is a temperature parameter. $E_{[CLS]}$ means the entire representation of a requirement. s(·) represents the cosine similarity of two vectors.


Based on the objective, our retriever learns the preference of LLMs, which will be effective to select helpful examples from the training set given a test requirement, resulting in promoting LLMs generating more correct programs.

\subsection{Inference}
\label{3.3}


During inference, instead of using heuristical approaches, we apply LAIL to select limited examples from the training set, which can retrieve examples with high metric scores given a test requirement and thus are helpful for LLMs in code generation. 
Specifically, we first feed requirements in the training set $\mathcal{R}$ to our trained retriever respectively and acquire their representational vectors $\left\{E_{CLS}^i\right\}_{i=1}^N$. Given a test requirement $x_t$, we obtain its representation $E_{CLS}^t$ by the retriever. Next, we match $E_{CLS}^t$ and $E_{CLS}^i$ and thus lead to $N$ pairs of representations $\left\{(E_{CLS}^t, E_{CLS}^i)\right\}_{i=1}^{N}$. We calculate their cosine similarities $\left\{c_i\right\}_{i=1}^N$ of all pairs and rank candidate examples according to similarity scores from high to low. Then, the top-$r$ examples are selected for ICL. We concatenate the top-$r$ examples as a prompt $\left\{e_1, \cdots, e_i, \cdots, e_r \right\}$ where their similarity scores gradually decrease (\eg, $c_1 < c_i < c_r$). We feed the prompt and the test requirement into LLMs and make LLMs generate programs with nuclear sampling \cite{holtzman2019curious} as described in Formula \ref{icl-formula}.

Note that LAIL only needs to encode training examples once. Given a test requirement, LAIL just calculates cosine similarities between it and all candidates. Thus, the efficiency of our approach is acceptable compared to other heuristic approaches.

\section{Study design}

To investigate the effectiveness of our LAIL, we perform a large-scale study to answer three research questions. In this section, we describe the details of our study, including datasets, evaluation metrics, baselines, base LLMs, and experimental details.

\subsection{Research Questions}


Our study aims to answer the following research questions.

\textbf{RQ1: How does LAIL perform compared to the state-of-the-art baselines?}
This RQ aims to verify that LAIL can generate more correct programs than state-of-the-art (SOTA) baselines. We apply three LLMs to evaluate our approach. We compare LAIL to 7 baselines in three code generation datasets and employ unit tests to check the correctness of generated programs.

\textbf{RQ2: Do developers prefer programs generated by LAIL?}
The ultimate goal of a code generation model is to assist developers in writing programs. In this RQ, we hire 10 developers to manually estimate the programs generated by our LAIL and baselines. We evaluate these programs in three aspects, including correctness, code quality, and maintainability.


\textbf{RQ3: What is the better design choice to estimate examples?}
In section \ref{3.1}, we leverage the prediction probability of ground-truth programs to estimate candidate examples. In this RQ, we explore other design choices to measure examples by LLMs themselves and compare them to our design.

\subsection{Datasets}
\label{datasets}

We conduct experiments on three code generation datasets, including MBJP \cite{athiwaratkun2022multi} in Java, MBPP \cite{austin2021program} in Python, and MBCPP \cite{athiwaratkun2022multi} in C++. The statistics of these datasets are given in Table \ref{tab:dataset}.

\begin{table}[t!]
  \centering
  \caption{Statistics of three datasets on the different split sets.}
    \begin{tabular}{lccc}
    \toprule
         & \multicolumn{1}{c}{MBJP} & \multicolumn{1}{c}{MBPP} & \multicolumn{1}{c}{MBCPP}  \\
    \midrule
    Language   &  Java & Python &  C++  \\
    \midrule
        \#Train   & 383  &  384 & 413   \\
        \#Dev  &90  & 90 & -- \\
      \#Test   & 493  &500 & 435   \\
      \midrule
    Avg. tests per example   &  3 & 3 &  3  \\
    Avg. tokens in requirement & 16.71 &16.50 &  17.38 \\
      Avg. tokens in code & 247.79  & 92.68 & 113.94   \\
    \bottomrule
    \end{tabular}%
    \vspace{-5mm}
  \label{tab:dataset}%
\end{table}%

\textbf{MBPP} \cite{austin2021program} contains 974 Python programming problems constructed by crowd-sourcing. Each example consists of a brief description, a single self-contained function solving the problem specified, and three test cases to evaluate the correctness of the generated programs. The problems range from simple numeric manipulations or tasks that require the basic usage of standard library functions to tasks that demand nontrivial external knowledge.

\textbf{MBJP} \cite{athiwaratkun2022multi}  and \textbf{MBCPP} \cite{athiwaratkun2022multi} are drived from MBPP \cite{austin2021program}. MBJP and MBCPP contain 966 and 848 crowd-sourced programming problems in Java and C++, respectively. Each problem consists of a description, an individual function, and three test cases. These problems cover programming fundamentals, standard library functionality, etc.

We follow previous studies \cite{athiwaratkun2022multi, austin2021program} to split three datasets into the training set, the valid set, and the test set, respectively. We measure the performance of different ICL approaches on the test set.

\subsection{Evaluation Metrics}
\label{metrics}



Following prior code generation studies \cite{chen2021evaluating, athiwaratkun2022multi}, we leverage Pass@k to evaluate our approaches. Pass@k evaluates the functional correctness of the generated programs by executing test cases. Precisely, given a test requirement, we generate $k$ programs using the sampling strategy. If any of the generated $k$ programs passes all test cases, we think the requirement is solved. Finally, the percentage of solved requirements in all test requirements is treated as Pass@k. In this paper, we set $k$ to 1, 3, and 5.

We notice that previous studies \cite{guo2020graphcodebert, feng2020codebert} also use some match-based metrics such as BLEU \cite{papineni2002bleu} and CodeBLEU \cite{ren2020codebleu}. These metrics focus on the textual similarity or syntactic similarity between reference programs and generated programs. However, existing work \cite{inala2022fault} has shown that match-based metrics can not effectively measure the functionality of programs. In this paper, we follow recent studies \cite{inala2022fault, radford2018improving, SCoT} use execution-based metrics (\eg Pass@k).

\subsection{Baselines}
\label{baselines}


There are a few studies to investigate in-context example selection for code generation such as zero-shot learning \cite{brown2020language}, random selection \cite{chen2021evaluating}, and AceCoder \cite{li2023towards}, where AceCoder \cite{li2023towards} is the state-of-the-art (SOTA) baseline.

\textbf{Zero-Shot Learning} \cite{brown2020language} directly inputs a requirement into LLMs without any examples as the prompt. LLMs generate programs for the given requirement.

\textbf{Random} \cite{chen2021evaluating} randomly select a few examples from the training set and construct a prompt. Then, LLMs predict the source code based on the prompt. 

\textbf{AceCoder}  \cite{li2023towards} uses BM25 \cite{robertson2009probabilistic} to calculate the textual similarities between a test requirement and the requirements of candidates, and retrieve a set of examples with high similarities. In inference, it first generates test cases and then predicts programs. For a fair comparison, we use the same pipeline to LAIL and all baselines.

To extensively evaluate the effectiveness of our LAIL, we also transfer some advanced ICL approaches in natural language processing to the source code.

\textbf{TOP-\normalsize{k}}-\textbf{SBERT} \cite{luo2023dr} leverages Sentence-BERT \cite{reimers2019sentence}, a representative sentence encode, to encode all requirements in the training set. Given a test requirement, we first encode it and computer semantic similarities between it and the training requirements. Next, we select the top-k similar examples.

\textbf{TOP-\normalsize{k}}-\textbf{GraphCodeBERT} \cite{guo2020graphcodebert} is a variant of TOP-\normalsize{k}-{SBERT}. It applies GraphCodeBERT to encode requirements and retrieve a few examples from the training set based on semantic similarity.  

\textbf{TOP-\normalsize{k}}-\textbf{VOTE} \cite{su2022selective} is a graph-based method \cite{su2022selective}, to votes examples. It first encodes each example by GraphCodeBERT and each example is as a vertice in the graph. Each vertice connects with its k nearest vertices based on their semantic similarities. Finally, it treats the  k nearest examples as a prompt.

\textbf{Uncertainty-Target} \cite{coleman2019selection} assumes examples with higher uncertainty have a greater impact on LLMs. It defines uncertainty as the perplexity when LLMs generate ground truths. The approach computes the uncertainty of each candidate and selects k items with high perplexity as a prompt.







\subsection{Base Large Language Models}
\label{llm}

This paper focuses on code generation with LLMs. Thus, we select three recently proposed LLMs for code generation as the base models. The details of the base models are shown as follows.

\textbf{ChatGPT} \cite{ChatGPT} is the state-of-the-art language model for code generation. It is trained on a large amount of natural language text and programming data. Then, ChatGPT is continually trained with reinforcement learning and learns to align with human instructions. We leverage OpenAI's APIs to access ChatGPT (\ie, gpt-3.5-turbo). 

\textbf{GPT-3.5} \cite{GPT} is a powerful large language model and is trained on large of unlabeled corpus. In this paper, we use OpenAI's APIs to access the latest version with 175 billion parameters (\ie, text-davinci-003).

\textbf{CodeGen} \cite{zhong2022codegen} is a family of language models for code generation. CodeGen is training with a 635GB code corpus and 1159GB English text data. In this paper, we leverage the largest version with 16 billion parameters (\ie CodeGen-Multi-16B).




\begin{table*}[htbp]
  \centering
  \caption{Evaluation results of our LAIL and baselines on CodeGen at the three code generation datasets. Numbers in bold indicate that the improvement is significant and the percentages in red color mean the relative improvements compared to the SOTA baseline.} 
    \begin{tabular}{lccccccccc}
    \toprule
          & \multicolumn{3}{c}{MBJP} & \multicolumn{3}{c}{MBPP} & \multicolumn{3}{c}{MBCPP}   \\
          \cmidrule(r){2-4} \cmidrule(r){5-7} \cmidrule(r){8-10} 
      & \multicolumn{1}{c}{Pass@1} & \multicolumn{1}{c}{Pass@3} & \multicolumn{1}{c}{Pass@5}  & \multicolumn{1}{c}{Pass@1} & \multicolumn{1}{c}{Pass@3} & \multicolumn{1}{c}{Pass@5} & \multicolumn{1}{c}{Pass@1} & \multicolumn{1}{c}{Pass@3} & \multicolumn{1}{c}{Pass@5}  \\
    \midrule
    Zero-Shot Learning \cite{brown2020language} &  14.27   & 23.69 & 27.83  & 8.80  &20.60  &25.60       &15.39  &25.97  & 30.94 \\
    Random \cite{chen2021evaluating} & 15.74 & 24.25 &28.14  &15.20  & 25.40 & 28.80 & 15.70 & 28.72  & 32.25 \\
    AceCoder \cite{li2023towards}   & 17.65  & 27.18  &30.63 &16.80  &26.40   & 29.40 & 17.47  & 30.11  &33.78  \\
\cmidrule{2-10}    TOP-\normalsize{k}-SBERT \cite{luo2023dr}   &16.63 &26.77  &29.21  &15.40  &25.00 &29.00 & 18.09  &29.43  & 33.33 \\
    TOP-\normalsize{k}-GraphCodeBERT \cite{guo2020graphcodebert}  & 17.44  & 24.34  & 28.40 & 17.40  &25.80 & 28.40  &18.16 &30.48 &35.47  \\
    TOP-\normalsize{k}-VOTE \cite{su2022selective}  &15.42 &21.70  &23.73   &17.40  &26.00 &29.40  &17.01 &30.03  & 35.63  \\
    Uncertainty-Target \cite{coleman2019selection}  & 17.09  & 23.06  & 27.93 & 14.60 & 24.20  &28.80  &17.70 &30.12 & 34.62 \\
    \cmidrule{2-10}   
    \cmidrule{2-10}    LAIL & \textbf{21.30}  & \textbf{28.49} & \textbf{32.05} & \textbf{18.60}  & \textbf{27.80} & \textbf{30.60} &\textbf{19.08}  &\textbf{31.36} &  \textbf{37.94} \\
    Relative Improvement  & {\color{red}{11.58\%}}  & {\color{red}{4.82\%}}  & {\color{red}{4.64\%}} & {\color{red}{6.89\%}} & {\color{red}{6.92\%}} & {\color{red}{4.08\%}} & {\color{red}{5.07\%}} &{\color{red}{2.89\%}} &{\color{red}{6.96\%}} \\
    \bottomrule
    \end{tabular}%
  \label{codegen}%
\end{table*}%

\subsection{Implementation Details}

\noindent \textbf{Estimate and Label Examples.}
We use greedy decoding to generate programs and collect the predicted probabilities of ground-truth programs. Following the previous generation works \cite{li2023towards}, we set the max generated length to 500 tokens. 
The number of examples in the set $S_i$ is 50 for efficiency. Note that the size of $S_i$ larger the better performance of LLMs and we leave other sizes in our future work.
We set $z$ and $v$ as 5 respectively and analyze their effects on performance in Section \ref{effect paprameters}. Thus, the positive set $S_i^p$ contains 5 examples with higher metric scores and the negative set $S_i^n$ has 5 items with lower scores.


\textbf{Training Neural Retriever.}
In this procedure, we set the number of the negative example set $\mathbb{N}_{i}$ as 64.
In other words, $h$ is set to 63. For each epoch, we randomly select 63 examples from the set $\mathcal{H}$. We also attempt to set the size of $\mathbb{N}_{i}$ to 32 and 128 in Section \ref{effect paprameters}. 
The learning rate is 5e-5 and the batch size is 32. We train the retriever for about 1 hour on 4 NVIDIA A100.

\textbf{Inference.} 
We treat a large language model as a black-box generator and sample programs from it.
The input of LLMs only contains a prompt and a test requirement without any natural language instructions. During the sampling, we use nuclear sampling \cite{holtzman2019curious} to decode, where the temperature is 0.8 and the top-p is 0.95. The maximum generated length is 500 tokens. For each test requirement, we generate 5 programs. For a fair comparison, we set the same parameters to generate programs for all baselines and our LAIL. 

\section{Results and Analysis}

\begin{table*}[htbp]
  \centering
  \caption{The performance of our LAIL and the existing approaches on GPT-3.5 at the three code generation datasets. The numbers in red color represent the relative improvements compared to the SOTA baseline.}
    \begin{tabular}{lccccccccc}
    \toprule
          & \multicolumn{3}{c}{MBJP} & \multicolumn{3}{c}{MBPP} & \multicolumn{3}{c}{MBCPP}   \\
          \cmidrule(r){2-4} \cmidrule(r){5-7} \cmidrule(r){8-10} 
      & \multicolumn{1}{c}{Pass@1} & \multicolumn{1}{c}{Pass@3} & \multicolumn{1}{c}{Pass@5}  & \multicolumn{1}{c}{Pass@1} & \multicolumn{1}{c}{Pass@3} & \multicolumn{1}{c}{Pass@5} & \multicolumn{1}{c}{Pass@1} & \multicolumn{1}{c}{Pass@3} & \multicolumn{1}{c}{Pass@5}  \\
    \midrule
    Zero-Shot Learning  \cite{brown2020language} & 44.83  & 53.05 & 59.72   & 20.00  & 27.00  & 29.60  &21.85 &37.94 &49.90  \\
    Random \cite{chen2021evaluating} & 47.87 & 59.83  &63.69  & 43.00  &56.40 & 60.80  &50.02 &61.43 & 65.28 \\
    AceCoder \cite{li2023towards}  & 50.91  &61.25 & 65.15 & 47.40 & 61.00 & 64.20  &53.25  &63.22  &66.21  \\
\cmidrule{2-10}    TOP-\normalsize{k}-SBERT \cite{luo2023dr}  & 50.30  & 60.46  & 64.70  & 48.00  & 61.40  & 64.00 &52.87 & 63.21 &67.13  \\
    TOP-\normalsize{k}-GraphCodeBERT \cite{guo2020graphcodebert} & 50.30  &60.85 & 64.50 & 49.20  &58.20  &64.20 & 52.64  &63.19  &67.28  \\
    TOP-\normalsize{k}-VOTE \cite{su2022selective} & 50.52  &60.45  & 63.49 & 47.80  &59.20  & 63.40  &51.03 &62.98   &65.51  \\
    Uncertainty-Target \cite{coleman2019selection}  &47.67  & 59.23  &63.69  & 37.80  &51.40  &57.80  & 42.87 &52.60  &57.18  \\
    \cmidrule{2-10}    
    \cmidrule{2-10}    LAIL  & \textbf{53.14}  & \textbf{62.87}  & \textbf{66.72} & \textbf{50.60} & \textbf{62.40} & \textbf{65.20}  & \textbf{54.71} &\textbf{65.98}  &\textbf{69.67}  \\ 
     Relative Improvement & {\color{red}{4.38\%}}  & {\color{red}{2.64\%}}  & {\color{red}{2.41\%}}  & {\color{red}{2.85\%}} & {\color{red}{1.63\%}} & {\color{red}{1.53\%}}  & {\color{red}{2.74\%}} &{\color{red}{4.37\%}} &{\color{red}{3.55\%}}  \\
    \bottomrule
    \end{tabular}%
  \label{codex}%
\end{table*}%

\noindent \textbf{RQ1: How does LAIL perform compared to the state-of-the-art baselines?}

In RQ1, we apply our LAIL and baselines to two LLMs and measure the correctness of generated programs.

\textbf{Setup.} We compare our LAIL and baselines (Section~\ref{baselines}) on three code generation datasets (Section~\ref{datasets}). The evaluation metric is Pass@k as described in Section~\ref{metrics}. For the metric, higher scores indicate better performance of approaches. 

\textbf{Results.} Table \ref{codegen} and Table \ref{codex} report the
Pass@k (k $\in$ [1, 3, 5]) of different approaches on CodeGen and GPT-3.5, respectively. Numbers in bold mean that the improvement is  significant compared with baselines, and the percentages in red color represent the relative improvements compared to the state-of-the-art (SOTA) baseline. 

\textbf{Analysis.} (1) \ul{LAIL achieves the best performance among all approaches with significant improvements.} 
In all datasets, LAIL generates more correct programs than baselines. Compared to the SOTA baseline, LAIL outperforms it by 11.58\%, 6.89\%, and 5.07\% on CodeGen, and acquires 4.38\%, 2.85\%, and 2.74\% improvement on GPT-3.5 in Pass@1, respectively. Note that Pass@1 is a very strict metric and is different to be improved. The significant improvements prove the superiority of our LAIL in code generation.
(2) \ul{Selecting proper examples is critical to the performance of ICL.}
Compared to random selection, LAIL acquires 8.31\%, 30.60\%, and 32.86\% absolute improvements in Pass@1 on GPT-3.5. AceCoder and other heuristic approaches further improve code generation performance by selecting textually or semantically similar examples. 
LAIL exploits LLMs themselves to estimate and trains a neural retriever to align with the preference of LLMs, then achieves the best results among all approaches. That demonstrates the importance of examples in ICL and verifies the reasonability of using LLMs themselves to label examples. 
(3)  \ul{LAIL is effective in LLMs with different sizes and different programming languages.} As above described, our approach on CodeGen and GPT-3.5 both achieve the best performance among all approaches. 
Table \ref{codegen} and Table \ref{codex} also show that LAIL can generate more correct programs on all datasets including MBPP (Python), MBJP (Java), and MBCPP (C++).
This reveals LAIL is generalized and can be applied to different LLMs and languages. 

\begin{tcolorbox}[enhanced,colback=gray!5!white,colframe=gray!75!black,drop fuzzy shadow southwest]
  \textbf{Answer to RQ1: LAIL achieves the best results among all baselines. In three datasets, LAIL acquires 11.58\%, 6.89\%,
and 5.07\% improvements on CodeGen, and achieves 4.38\%, 2.85\%, and 2.74\% improvements on GPT-3.5 at Pass@1. The significant improvements prove our approach can align with the preference of LLMs. Thus, it is able to select suitable examples for ICL and generates more correct programs.} 
\end{tcolorbox}

\begin{table}[htbp]
  \centering
  \caption{The results of human evaluation. The numbers in red represent LAIL' relative improvements compared to the SOTA baseline. ``CR'', ``QL'', and ``MT'' mean correctness, code quality, and maintainability, respectively.}
    \begin{tabular}{lccc}
    \toprule
    Approach  & \multicolumn{1}{c}{CR} & \multicolumn{1}{c}{ QL} & \multicolumn{1}{c}{MT } \\
    \midrule
    Zero-Shot Learning \cite{brown2020language} & 0.472  &1.079 & 1.183 \\
    Random \cite{chen2021evaluating}  & 0.925  &  1.416 &  1.520  \\
    AceCoder \cite{li2023towards} & 1.523  & 1.652  & 1.647 \\
    TOP-\normalsize{k}-SBERT \cite{luo2023dr} & 1.569  & 1.640  & 1.665 \\
    TOP-\normalsize{k}-GraphCodeBERT \cite{guo2020graphcodebert}  & 1.582  & 1.738  & 1.609\\
    TOP-\normalsize{k}-VOTE \cite{su2022selective}  & 1.325  & 1.624  &1.631  \\
     Uncertainty-Target \cite{coleman2019selection} & 0.736  &1.297   &1.324  \\
      \cmidrule{2-4}
    LAIL & \textbf{1.764}  & \textbf{1.839} & \textbf{1.782} \\ 
    Relative Improvement  &  {\color{red}{11.504\%}}  & {\color{red}{5.811\%}}   & {\color{red}{7.027\%}} \\
    \bottomrule
    \end{tabular}%
  \label{human}%
\end{table}%

\noindent \textbf{RQ2: Do developers prefer programs generated by LAIL?}

The goal of a code generation approach is to assist developers in writing programs. Thus, a good program not only satisfies the requirement but also is easy to read and maintain. In this question, we manually verify generated programs in three aspects.

\textbf{Setup.} 
Following previous work \cite{SCoT}, we manually measure programs generated by different approaches in three aspects (\eg  correctness, code quality, maintainability). Correctness measures whether a program satisfies the given requirement, code quality verifies whether a program does not contain bas code smell, and maintainability measures whether the implementation is standardized and easy to read. For each aspect, the score is an integer and ranges from 0 to 2. The higher score, the better code is.

Specifically, we randomly select 50 test samples from MBPP, MBJP, and MBCPP, respectively \footnote{The confidence level is 85\% and the margin of error is 15\% with 50 selected examples.}. Then, we use LAIL and baselines to generate programs for these examples. Finally, we obtain 400 (50 $\times$ 8) programs for human evaluation. The 400 programs are randomly divided into 5 groups. We hire 10 developers to verify these programs. These developers are computer science students or industrial practitioners and write programs at least for 3 years. Each group is measured by two developers and the final score is the average of two developers' scores. 

\textbf{Results.}  The results of the human evaluation are shown in Table \ref{human}. The percentages in parentheses are the relative improvements compared to the SOTA baseline. 

\textbf{Analyses.}
(1) \ul{Developers prefer programs generated by LAIL over all baselines.} LAIL substantially outperforms all baselines in three aspects, which demonstrates the effectiveness of our approach. (2) \ul{In addition to satisfying requirements, programs provided by our approach are easier to read and have less bad smell.}
Particularly, our LAIL surpasses the SOTA approach by 5.811\% in code quality and 7.027\% in maintainability. (3) \ul{Some heuristic methods that select the same examples as prompts for all test requirements are not optimal for ICL.} As shown in Figure \ref{human}, Uncertainty-Target and Vote-K perform comparably to the random approach, which indicates that applying the same prompt to all test data is not a good choice, and different test requirements should use their appropriate prompts. 

\begin{tcolorbox}[enhanced,colback=gray!5!white,colframe=gray!75!black,drop fuzzy shadow southwest]
  \textbf{Answer to RQ2: Human evaluation shows that LAIL significantly outperforms SOTA baselines on all three aspects. Programs generated by LAIL can better satisfy requirements, be easier to read, and have little bad code smell. That is, developers prefer programs generated by LAIL among all approaches.} 
\end{tcolorbox}

\noindent \textbf{RQ3: What is the better design choice to estimate examples?}

In this paper, we apply the predicted probability of ground truths to estimate examples. We further investigate two plausible choices to measure candidates by LLMs themselves.

\textbf{Setup.} 
We design two plausible approaches including the Match-BLEU and the Match-CodeBLEU formats. 
The Match-BLEU format uses the BLEU score of the generated program to measure examples. The Match-CodeBLEU approach applies the CodeBLEU score of the predicted program to label examples. In this RQ, we select a representative LLM (\ie  GPT-3.5) as the base model and evaluate their performance on three datasets.

\textbf{Results.} The results of different feedback scores are represented in Table \ref{feedback}. For comparison, we also present the results of the existing SOTA baseline in Table \ref{feedback}.

\textbf{Analyses.} (1) \ul{Our probability-based approach is better than Match-BLEU and Match-CodeBLEU.}  In all datasets, our LAIL outperforms them by 4.05\%, 5.86\%, and 3.77\% on three datasets at Pass@1, respectively. We argue that the predicted probability of ground truth accurately reflects the certainty of LLMs for correct programs. The BLEU-based and CodeBLEU-based methods only reflect the literal accuracy of LLMs' prediction, but can not provide how certain LLMs are when predicting programs. 
(2) \ul{Match-CodeBLEU approach is more effective than Match-BLEU method.}  Match-CodeBLEU outperforms Match-BLEU on all datasets. The reason might be that CodeBLEU measures the n-gram match, the semantic match, and the syntactic match between the hypothesis programs and the reference code tokens, which is better to evaluate the generated programs than BLEU.

\begin{tcolorbox}[enhanced,colback=gray!5!white,colframe=gray!75!black,drop fuzzy shadow southwest]
  \textbf{Answer to RQ3: Probability-based approach is better than Match-CodeBLEU and Match-BLEU formats. It can effectively reflect the LLMs' preference for candidate examples given a test requirement. .} 
\end{tcolorbox}

\begin{table*}[htbp]
  \centering
  \caption{The comparison of different formats to estimate examples on three  datasets. The values in parentheses mean significant improvements. ``Probability-Based '' represents our LAIL.}
    \begin{tabular}{lccccccccc}
    \toprule
          & \multicolumn{3}{c}{MBJP} & \multicolumn{3}{c}{MBPP} & \multicolumn{3}{c}{MBCPP}   \\
          \cmidrule(r){2-4} \cmidrule(r){5-7} \cmidrule(r){8-10} 
      & \multicolumn{1}{c}{Pass@1} & \multicolumn{1}{c}{Pass@3} & \multicolumn{1}{c}{Pass@5}  & \multicolumn{1}{c}{Pass@1} & \multicolumn{1}{c}{Pass@3} & \multicolumn{1}{c}{Pass@5} & \multicolumn{1}{c}{Pass@1} & \multicolumn{1}{c}{Pass@3} & \multicolumn{1}{c}{Pass@5}  \\
    \midrule
    Random \cite{chen2021evaluating}  &47.87  &59.83   &63.69 &43.00 &56.40  &60.80  &50.04  &61.45  & 65.28 \\
    Match-BLEU  & 50.19 &59.87 & 63.92   & 46.20 & 60.20 & 62.40 & 50.26  & 61.47  &65.73  \\
   Match-CodeBLEU  & 51.07 & 60.04  & 64.58  & 47.80 &59.20 & 63.80  & 52.72  &62.03 & 66.80 \\
   Probability-Based (LAIL)  & \textbf{53.14}  & \textbf{62.87}  & \textbf{66.72}  & \textbf{50.60}  & \textbf{62.40}  & \textbf{65.20}  & \textbf{54.71} & \textbf{65.98}  & \textbf{69.67}  \\
   \cmidrule{2-10}  Relative Improvement  & {\color{red}{4.05\%}}  & {\color{red}{4.71\%}}  & {\color{red}{3.31\%}} & {\color{red}{5.86\%}} & {\color{red}{3.65\%}} & {\color{red}{2.19\%}} & {\color{red}{3.77\%}} &{\color{red}{6.37\%}} &{\color{red}{4.30\%}}  \\
    \bottomrule
    \end{tabular}%
  \label{feedback}%
\end{table*}%

\section{Discussion}

\subsection{Transferability}

We explore whether our retriever based on one LLM's feedback and specific dataset can be transferred to other LLMs or code generation datasets without further tuning. This is a significant research question since the retriever for each LLM and dataset needs to be trained in real applications.

\subsubsection{Transfer across LLMs}

We consider transferring the retriever based on one LLM's feedback to another LLM. Specifically, we use a source LLM (\eg CodeGen or GPT-3.5) to estimate examples for training  a retriever and then apply the retriever to another target LLM (\eg ChaTGPT) in generating programs. Table \ref{transfer llms} shows the performance of ChaTGPT in three datasets. We surprisingly find our retriever based on CodeGen and GPT-3.5 can bring obvious improvements to ChaTGPT. In particular, in terms of Pass@1, ChaTGPT achieves 2.56\% improvements from CodeGen’s feedback and 4.31\% enhancements from GPT-3.5’s feedback compared to the SOTA baseline. The phenomenons demonstrate that our approach has satisfying transfer ability across different LLMs. Note that ChaTGPT can not provide the prediction probability of ground truths in practice, thus LAIL is a quite meaningful approach, especially for LLMs whose parameters are unavailable.
Besides, the performance of ChaTGPT from GPT-3.5’s feedback is higher than the counterpart from CodeGen’s feedback. The reason might be that GPT-3.5 and ChatGPT have comparable abilities in code generation, thus their preference is similar and GPT-3.5 can provide more proper examples as prompts.

To verify the transfer ability among LLMs with different sizes, we further evaluate the performance of CodeGen based on GPT-3.5's feedback and the results of GPT-3.5 from CodeGen's feedback, where the parameter size of CodeGen is much smaller than the counterpart of GPT-3.5. As shown in Table \ref{transfer llms}, compared to the random approach, the retriever learned from CodeGen achieves 8.25\%, 12.09\%, and 4.32\% relative improvements to GPT-3.5 on Pass@1, meanwhile our retriever under GPT-3.5's feedback brings  22.30\%, 14.47\%, and 16.94\% improvements to CodeGen, respectively. This indicates our approach has satisfying  transfer ability among LLMs and can bring improvements to target LLMs.


\subsubsection{Transfer across datasets}

\begin{figure}[t]
\centering
\includegraphics[width=0.8\linewidth]{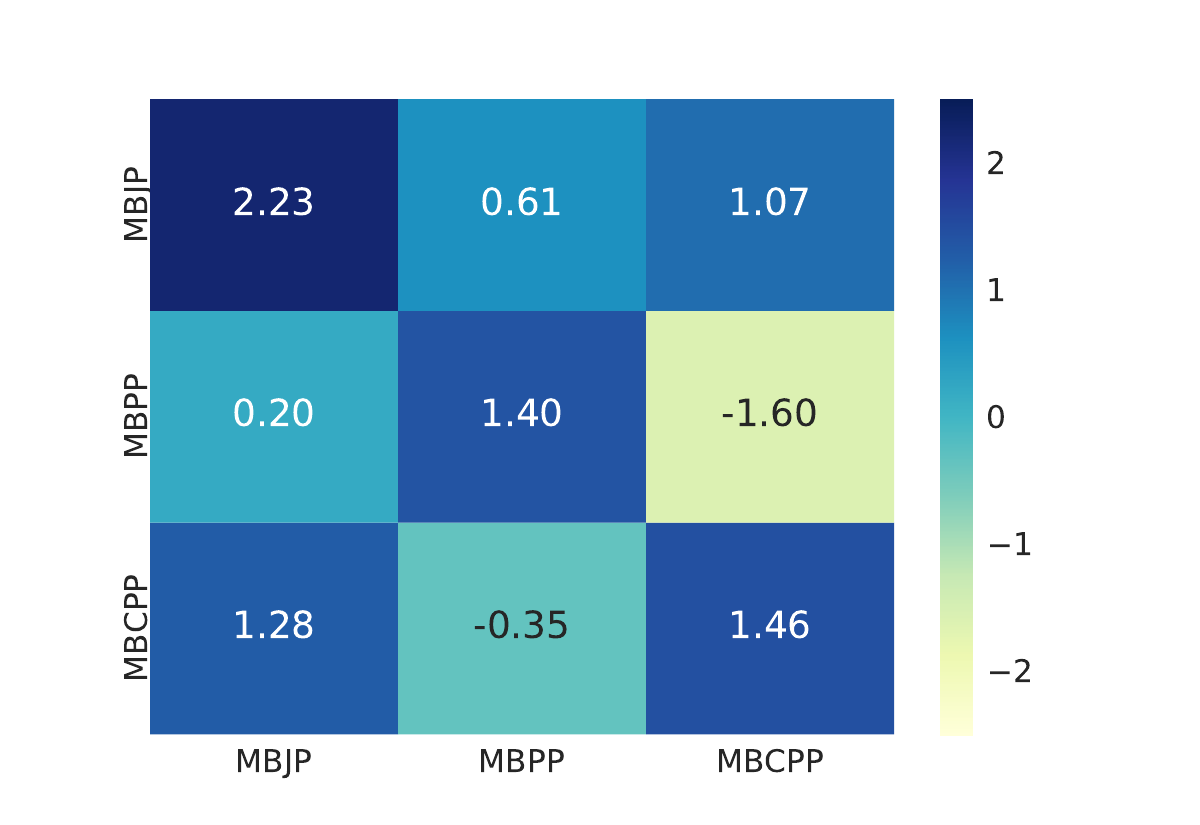}
\caption{Results of transferring the retriever trained on one dataset (row) to others (column) on GPT-3.5 in MBPP dataset.}
\label{cross datasets}
\end{figure}

Considering that the compositional features of natural language are general, the retriever trained on one dataset may apply to other datasets and exploit similar knowledge in different datasets. In this section, we further investigate whether a retriever trained on one dataset can transfer to others. We transfer the retriever among three datasets (\eg, MBJP, MBPP, and MBCPP) and Figure \ref{cross datasets} demonstrates the transferring results on Pass@1. We find that most retrievers can successfully transfer to other datasets and bring improvements compared to their SOTA baselines. Concretely, the retriever trained on MBJP (MBCPP) achieves 1.28\% (1.07\%) absolute improvements when it migrates to MBCPP (MBJP). Meanwhile, the MBCPP-trained retriever hardly transfers to MBPP and the MBPP-based retriever suffers the generation performance on MBPP. The reason might be that Java and C++ are object-oriented programming languages, and their syntax and code morphology are similar.
Exploring a retriever that is suitable for many datasets is a challenging but meaningful research question, and we leave this topic as our future work.

\begin{table*}[htbp]
  \centering
  \caption{Results of transferring a retriever learned on one LLM to others on three datasets. $\clubsuit$ means the source LLM.}
    \begin{tabular}{lccccccccc}
    \toprule
        \multicolumn{1}{l}{\multirow{2}[4]{*}{$\clubsuit$ ChaTGPT}}  & \multicolumn{3}{c}{MBJP} & \multicolumn{3}{c}{MBPP} & \multicolumn{3}{c}{MBCPP}   \\
           \cmidrule(r){2-4} \cmidrule(r){5-7} \cmidrule(r){8-10} 
      & \multicolumn{1}{c}{Pass@1} & \multicolumn{1}{c}{Pass@3} & \multicolumn{1}{c}{Pass@5}  & \multicolumn{1}{c}{Pass@1} & \multicolumn{1}{c}{Pass@3} & \multicolumn{1}{c}{Pass@5} & \multicolumn{1}{c}{Pass@1} & \multicolumn{1}{c}{Pass@3} & \multicolumn{1}{c}{Pass@5}  \\
    \midrule
    Zero-Shot Learning \cite{brown2020language} &16.63  &34.48   &44.21 & 26.60  & 32.00  &34.60   &30.34  & 57.01  & 63.68 \\
    Random \cite{chen2021evaluating}  & 53.34 & 62.27 &65.72 &50.80 &60.60 & 63.40 & 40.15 &60.06 &66.28  \\
    AceCoder \cite{li2023towards}  & 54.46 &64.01 & 66.75 &54.40 &62.40  & 65.20 & 42.53 & 62.29 &68.32 \\
\cmidrule{2-10}   TOP-\normalsize{k}-SBERT \cite{luo2023dr} & 54.26 & 63.89 &66.53 & 54.80 & 63.00 &65.40 & 43.37  &61.45 & 67.11 \\
    TOP-\normalsize{k}-GraphCodeBERT \cite{guo2020graphcodebert}  &53.95 & 62.67  & 66.32 & 54.20 & 62.60 &65.20 & 44.08 &61.61 & 68.27 \\
    TOP-\normalsize{k}-VOTE \cite{su2022selective}  & 51.93   & 62.88 & 65.72 & 42.00 & 56.00  &61.00 &42.74 &62.68 & 67.73 \\
    Uncertainty-Target \cite{coleman2019selection} & 49.69 & 60.45  & 54.50 &36.40  & 51.80 &56.80  & 42.46  &61.84   &66.67  \\
        \midrule
     & \multicolumn{9}{c}{CodeGen} \\
    \cmidrule{2-10}    LAIL   & \textbf{54.79}  & \textbf{64.70}  & \textbf{67.43} & \textbf{55.60}  & \textbf{63.80}  & \textbf{66.20} & \textbf{45.21}  & \textbf{63.45}  & \textbf{68.93} \\
    \cmidrule{2-10}  Relative Improvement  & {\color{red}{0.61\%}}  & {\color{red}{1.08\%}}  & {\color{red}{1.02\%}} & {\color{red}{1.44\%}} & {\color{red}{1.27\%}} & {\color{red}{1.22\%}}  & {\color{red}{2.56\%}} &{\color{red}{1.23\%}} &{\color{red}{0.89\%}} \\
    \midrule
    & \multicolumn{9}{c}{GPT-3.5} \\
    \cmidrule{2-10}    LAIL  & \textbf{55.97}  & \textbf{64.97}  & \textbf{68.27}   & \textbf{56.20} & \textbf{64.60} & \textbf{66.80}  & \textbf{45.98}  & \textbf{63.84} & \textbf{70.35}  \\ 
    \cmidrule{2-10}  Relative Improvement  & {\color{red}{2.77\%}}  & {\color{red}{1.50\%}}  & {\color{red}{2.28\%}} & {\color{red}{2.56\%}} & {\color{red}{2.54\%}} & {\color{red}{2.14\%}} & {\color{red}{4.31\%}} &{\color{red}{1.85\%}} &{\color{red}{2.97\%}} \\
    \midrule
    \midrule
   \multicolumn{1}{l}{$\clubsuit$ GPT-3.5}  & \multicolumn{9}{c}{CodeGen} \\
   \cmidrule{2-10}    Random \cite{chen2021evaluating}  & 47.87  & 59.83  & 63.69 & 43.00 & 56.40 & 60.80  &50.02  & 61.43 & 65.28  \\ 
    \cmidrule{2-10}    LAIL  & \textbf{51.82}  & \textbf{61.75}  & \textbf{64.89} & \textbf{48.20} & \textbf{59.00} & \textbf{64.20}   & \textbf{52.18}  & \textbf{64.54} & \textbf{67.90}  \\ 
    \cmidrule{2-10}  Relative Improvement  & {\color{red}{8.25\%}}  & {\color{red}{3.21\%}}  & {\color{red}{1.88\%}} & {\color{red}{12.09\%}} & {\color{red}{4.61\%}} & {\color{red}{5.59\%}}  & {\color{red}{4.32\%}} &{\color{red}{5.06\%}} &{\color{red}{4.01\%}} \\
    \midrule
    \midrule
     \multicolumn{1}{l}{$\clubsuit$ CodeGen} &  \multicolumn{9}{c}{GPT-3.5} \\
     \cmidrule{2-10}    Random  \cite{chen2021evaluating}  & 15.74 & 24.25 & 28.14 & 15.20 & 25.40 &28.80  &15.70  & 28.72 & 32.25  \\ 
    \cmidrule{2-10}    LAIL   & \textbf{19.25}  & \textbf{27.53}  & \textbf{30.98} & \textbf{17.40} & \textbf{26.00} & \textbf{30.00}  & \textbf{18.36}  & \textbf{30.41} & \textbf{35.54}  \\ 
    \cmidrule{2-10}  Relative Improvement  & {\color{red}{22.30\%}}  & {\color{red}{13.53\%}}  & {\color{red}{10.09\%}} & {\color{red}{14.47\%}} & {\color{red}{2.36\%}} & {\color{red}{4.17\%}}  & {\color{red}{16.94\%}} &{\color{red}{5.88\%}} &{\color{red}{10.20\%}} \\
    \bottomrule
    \end{tabular}%
  \label{transfer llms}%
\end{table*}%

\subsection{Impacts of In-Context Example Numbers}

\begin{figure}[t]
\centering
\includegraphics[width=0.8\linewidth]{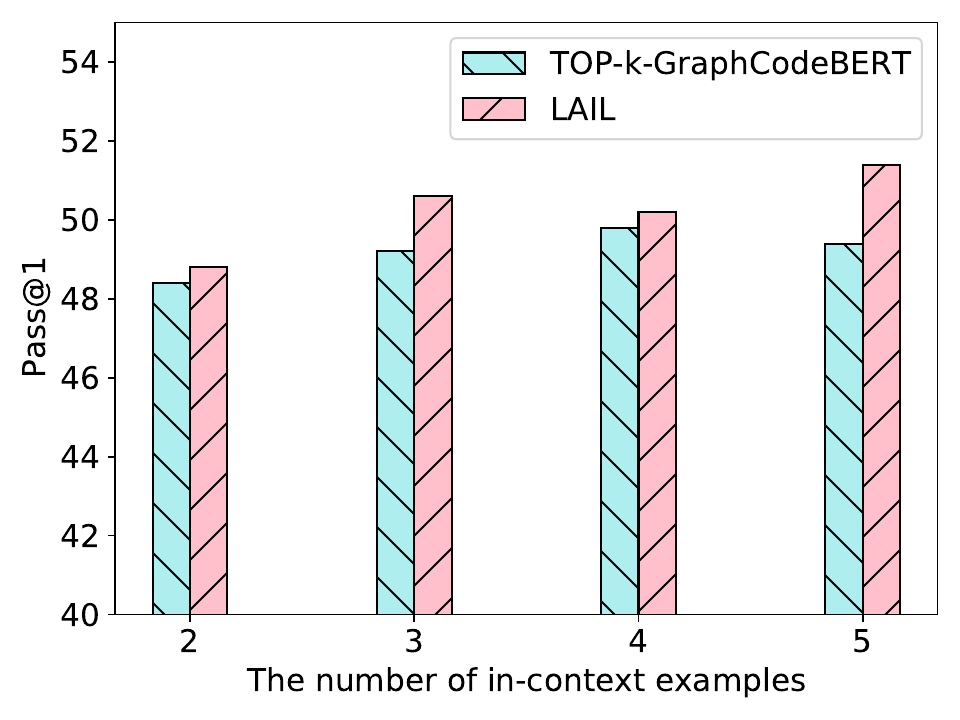}
\caption{The performance of the different number of in-context examples on GPT-3.5 in MBPP datasets.}
\label{number}
\end{figure}

Most of the LLMs are trained with a limited input length, which restricts the number of examples in the prompt. Previous studies \cite{luo2023dr, dong2022survey} also find that LLMs are affected by the number of in-context examples. Here we explore the impacts of the number of examples in the prompt on baselines and our approach. Figure \ref{number} reports how the performance of LAIL and GraphCodeBET change with respect to different in-context example numbers inMBPP dataset. Note that GraphCodeBET is the best baseline in this dataset, thus we choose it to compare with our approach.
We observe that the performances of GraphCodeBET and our approach monotonically increase in rough with the increase of in-context example numbers. In addition, our LAIL always outperforms GraphCodeBERT in all cases, which further proves the superiority of our approach even with different in-context example numbers.

\subsection{Effects of Contrastive Learning Parameters}
\label{effect paprameters}

The number of negative examples $\mathbb{N}_{i}$ (named $\mathbb{N}_{i}$ for convenience in Figure \ref{parameters}) is an important element in contrastive learning for training a neural retriever. We investigate how the element affects the performance of our approach. As shown in Figure \ref{parameters}, with the increasing of negative example numbers, LAIL achieves consistent improvements and can generate more correct programs. 
Besides, as described in \ref{3.2}, we randomly select an example $e_i^n$ from $S_i^n$ into $\mathbb{N}_{i}$. We also explore the effect of the number of selected examples from $S_i^n$ (dubbed $\tau_{ne}$). We can find that the more hard negative examples, the worse our approach performs. We argue that the examples of $S_i^n$ are with high BM25 scores among all candidate examples and thus $S_i^n$ may contain some false negative examples. Selecting more examples from $S_i^n$ will affect a retriever learning for choosing suitable in-context examples.
Between the two factors, the number of selected examples from $S_i^n$ has a greater influence on the performance of our approach.

\begin{table}[t]
  \centering
  \caption{Effects of parameters.}
    \begin{tabular}{cccc}
    \toprule
   & \multicolumn{1}{c}{Pass@1} & \multicolumn{1}{c}{Pass@3} & \multicolumn{1}{c}{Pass@5} \\
    \midrule
    Negative examples ($\mathbb{N}_{i}$) &  &   &  \\
    32    & 50.20 & 61.80 & 65.00 \\
    64    &50.60  &62.40 & 65.20  \\
    128   & 50.80 &62.40 & 65.40 \\
    \midrule
    $\tau_{ne}$ &       &  \\
    1     &50.60  &62.40 & 65.20 \\
    5     &50.20  & 62.20 & 65.00   \\
    10     &49.40  & 61.80& 64.40\\
    \bottomrule
    \end{tabular}%
    \vspace{-6mm}
  \label{parameters}%
\end{table}%

\subsection{Threats to Validity}

There are two main threats to the validity of our work. 

\textbf{The generalizability of our experimental results.} 
For the datasets, we follow previous studies \cite{chen2021evaluating, austin2021program, athiwaratkun2022multi} and leverage three representative code generation datasets. The three datasets cover different programming languages (\eg, Java, Python, and C++) and come from real-world software communities. To verify the superiority of LAIL, we consider seven existing ICL approaches in both the code generation task and many maintain natural language tasks. In addition, to effectively evaluate our approach, we select a series of advanced pre-trained large language models (CodeGen \cite{zhong2022codegen}, GPT-3.5, and ChatGPT) as base models in the past three years. We apply our approach and baselines to base models and evaluate their performance in code generation. For the metric, following existing works \cite{chen2022codet, zhong2022codegen}, we select a widely used Pass@k metric to evaluate all approaches. It is an execution-based metric that utilizes test cases to check the correctness of generated programs. To ensure fairness, we execute each method three times and report the average experimental results.

\textbf{The implementation of models and prompts.}
It is widely known that deep neural models are sensitive to the implementation details. In this paper, we need to execute all baselines and our approach on three base models. For baselines, we apply the source code and parameters published by their original papers \cite{guo2020graphcodebert, zhong2022codegen, reimers2019sentence}. For LLMs (\eg, GPT-3.5 and ChatGPT), the hyper-parameters (\eg, temperature) of sampling will impact their outputs. In experiments, we follow previous studies \cite{zhong2022codegen, SCoT} to set hyper-parameters for all approaches. 
In addition, the performance of LLMs heavily depends on the prompts, including the instruction and the number of examples. To alleviate this threat, we leverage the same number of examples for all approaches and directly construct prompts without any natural language instructions.
Besides, it's worth noting that the candidate pool for example selection also influences ICL performance. A large-scale study on 13.2 million real code files showed the proportion of reused code is up to 80\% \cite{mockus2007large}. Thus, we use the training set of each dataset as the candidate set for all approaches in this paper and believe that the training set can provide supporting examples for test requirements. We do not tune the prompt and hyper-parameters experimentally and set them empirically. Thus, there might be room to tune hyper-parameter settings of our approach for more improvements.

\section{Related Work}

\subsection{Pre-trained Language Models}

Code generation aims to automatically generate programs given natural language requirements. Pre-trained language models have achieved state-of-the-art results in code generation. They are pre-trained on large-scale unlabeled code corpus and then transferred to code generation. Existing pre-trained models can be divided into encoder-decoder models and decoder-only models.

\textbf{Encoder-decoder models} consist of an encoder and a decoder. An encoder takes a requirement as the input, and a decoder outputs a program. Many popular encoder-decoder architectures in natural language processing have been applied to source code, which often use denoising-based pre-training objectives to pre-train, such as masked sequence prediction, identifier tagging, and masked identifier prediction. For example, CodeT5 \cite{wang2021codet5} is the variant of T5 model \cite{T5} to support programs. \cite{PLBART} is adapted from the BART model \cite{lewis2019bart}, which is pre-trained on a number of programs. Besides, some studies \cite{chakraborty2022natgen} further consider the naturalized feature of code and train models to generate natural programs based on noised programs.

\textbf{Decoder-only models} contain decoder networks, which are pre-trained with the next token prediction objective. Inspired by the success of GPT series \cite{radford2018improving} in natural language processing, researchers attempt to adapt similar models to source code.
CodeGPT \cite{lu2021codexglue} is pre-trained on CodeSearchNet \cite{husain2019codesearchnet}  corpus with the same setting and structure as GPT-2. CodeX is continually fine-tuned on GPT-3 \cite{brown2020language} on code corpus from GitHub.  CodeX is proficient in over a dozen languages (\eg, JavaScript and Python) and supports a commercial application (\eg, Copilot).
Since its parameters are not available, many researchers try to replicate them and bring CodeParrot \cite{CodeParrot}, GPT-CC \cite{GPT-CC}, PyCodeGPT \cite{zan2022cert} and CodeGen \cite{zhong2022codegen}. CodeGeeX \cite{zheng2023codegeex} is pre-trained on a large code corpus and can generate more than 20 programming languages. Lately, ChatGPT is proposed by OpenAI and achieves impressive performance in code generation. In this paper, we select three representative networks including CodeGen \cite{zhong2022codegen}, GPT-3.5 \cite{GPT}, and ChatGPT \cite{ChatGPT} as our base models.

\subsection{In-Context Learning}

In-context learning (ICL) is an emerging approach to using large language models (LLMs). By providing limited examples as a prompt, ICL empowers LLMs to learn a specific downstream task.
ICL \cite{brown2020language} is first proposed in natural language processing and achieves impressive performance in many tasks \cite{levy2022diverse, dong2022survey} such as text generation and sentiment classification.

Inspired by the success of ICL in natural language processing, researchers attempt to adapt ICL to source code \cite{bareiss2022code, chen2021evaluating, chen2023teaching}. They design task-specific instructions with a set of examples to prompt LLMs and improve the performance on many tasks (\eg, code generation \cite{chen2023teaching} and code repair \cite{jiang2023impact}).
The popularity of ICL also introduces instability in performance: given different sets of examples as prompts, LLMs' performance usually varies from random to near state-of-the-art \cite{liu2021makes}. Some studies \cite{chen2023teaching, li2023towards} make efforts to mitigate this issue. Researchers \cite{chen2023teaching} randomly select limited examples for ICL and verify the results in code generation. AceCoder \cite{li2023towards} uses BM25 to select n-gram matching examples and further generates more correct programs. However, these approaches are based on lexical features, which require developers to design heuristics and lead to sub-optimal performance.
Compared to the existing methods, LAIL is a learning-based approach for selecting in-context examples. Instead of considering textual similarity, it applies LLMs themselves to measure candidate examples and uses contrastive learning to learn LLMs' preferences in code generation.

\section{Conclusion}

Due to the instability of ICL, there is an increasing demand for in-context example selection in code generation. This paper proposes a novel learning-based ICL approach LAIL. We exploit LLMs themselves to estimate examples by considering the generation probabilities of ground-truth programs given a requirement, and label these examples based on LLMs' feedback. We then introduce a contrastive learning objective to train a retriever, resulting in acquiring the preferences of LLMs. Experimental results on three code generation datasets demonstrate that our proposed approach achieves excellent performance. LAIL also shows satisfactory transferability across LLMs and datasets, showing that it is an efficient and practical approach to code generation. In the future, we will apply our approach to other LLMs and datasets.



\bibliographystyle{FSE23}
\bibliography{FSE23}



\end{document}